\documentclass[]{spie}  
\usepackage{graphicx}
\usepackage{rotating}
\usepackage{amssymb}
\usepackage{mathptmx}
\usepackage{epstopdf}
\usepackage[]{natbib}
\usepackage{lineno}

\bibpunct{[}{]}{,}{n}{}{,}

\begin{document}

\newcommand{\hdblarrow}{H\makebox[0.9ex][l]{$\downdownarrows$}-}
\newcommand{\pb}{\protect\textsc{polarbear}}
\newcommand{\Pb}{\protect\textsc{Polarbear}}

\title{The \Pb-2 and the Simons Array Experiments}

\author{A. Suzuki$^{a, b}$ ,  
P. Ade$^{d}$ ,  
Y. Akiba$^{e,x}$ ,  
C. Aleman$^{f}$ ,  
K. Arnold$^{y}$ , 
C. Baccigalupi$^{g}$ , 
B. Barch$^{a}$ , 
D. Barron$^{a}$ , 
A. Bender$^{h}$ , 
D. Boettger$^{m}$ , 
J. Borrill$^{i}$ , 
S. Chapman$^{j}$ , 
Y. Chinone$^{a}$ , 
A. Cukierman$^{a}$ , 
M. Dobbs$^{k}$ , 
A. Ducout$^{l}$ , 
R. Dunner$^{m}$ , 
T. Elleflot$^{f}$ , 
J. Errard$^{1}$ , 
G. Fabbian$^{g}$ , 
S. Feeney$^{l}$ , 
C. Feng$^{n}$ , 
T. Fujino$^{c}$ , 
G. Fuller$^{f}$ , 
A. Gilbert$^{k}$ , 
N. Goeckner-Wald$^{a}$ , 
J. Groh$^{a}$ , 
T. De Haan$^{a}$ , 
G. Hall$^{a}$ , 
N. Halverson$^{o}$ , 
T. Hamada$^{e}$ , 
M. Hasegawa$^{e}$ , 
K. Hattori$^{e}$ , 
M. Hazumi$^{c,e,x}$ , 
C. Hill$^{a}$ , 
W. Holzapfel$^{a}$ , 
Y. Hori$^{a}$ , 
L. Howe$^{f}$ , 
Y. Inoue$^{e,2}$ , 
F. Irie$^{c}$ , 
G. Jaehnig$^{o}$ , 
A. Jaffe$^{l}$ , 
O. Jeong$^{a}$ , 
N. Katayama$^{c}$ , 
J. Kaufman$^{f}$ , 
K. Kazemzadeh$^{f}$ , 
B. Keating $^{f}$ , 
Z. Kermish$^{p}$ , 
R. Keskitalo$^{i}$ , 
T. Kisner$^{i}$ , 
A. Kusaka$^{q}$ , 
M. Le Jeune$^{r}$ , 
A. Lee$^{a}$ , 
D. Leon$^{f}$ , 
E. Linder$^{q}$ , 
L. Lowry$^{f}$ , 
F. Matsuda$^{f}$ , 
T. Matsumura$^{s}$ , 
N. Miller$^{t}$ , 
K. Mizukami$^{c}$ , 
J. Montgomery$^{k}$ , 
M. Navaroli$^{f}$ , 
H. Nishino$^{e}$ , 
J. Peloton$^{r}$ , 
D. Poletti$^{r}$ , 
G. Rebeiz$^{u}$ , 
C. Raum$^{a}$ , 
C. Reichardt$^{v}$ , 
P. Richards$^{a}$ , 
C. Ross$^{j}$ , 
K. Rotermund$^{j}$ , 
Y. Segawa$^{e}$ , 
B. Sherwin$^{q}$ , 
I. Shirley$^{a}$ , 
P. Siritanasak$^{f}$ , 
N. Stebor$^{f}$ , 
R. Stompor$^{r}$ , 
J. Suzuki$^{e}$ , 
O. Tajima$^{e}$ , 
S. Takada$^{w}$ , 
S. Takakura$^{e,z}$ , 
S. Takatori$^{e}$ , 
A. Tikhomirov$^{j}$ , 
T. Tomaru$^{e}$ , 
B. Westbrook$^{a}$ , 
N. Whitehorn$^{a}$ , 
T. Yamashita$^{c}$ , 
A. Zahn$^{f}$ , 
O. Zahn$^{a}$
\skiplinehalf
\small{
$^{a}$ Department of Physics, University of California, Berkeley, CA 94720, USA
\\$^{b}$ Radio Astronomy Laboratory, University of California, Berkeley, CA 94720, USA
\\$^{c}$ Kavli IPMU (WPI), UTIAS, The University of Tokyo, Kashiwa, Chiba 277-8583, Japan
\\$^{d}$ School of Physics and Astronomy, Cardiff University, Cardiff CF10 3XQ, United Kingdom
\\ $^{e}$ High Energy Accelerator Research Organization (KEK), Tsukuba, Ibaraki 305-0801, Japan
\\ $^{f}$ Department of Physics, University of California, San Diego, CA 92093-0424, USA
\\$^{g}$ International School for Advanced Studies (SISSA), Via Bonomea 265, 34136, Trieste, Italy
\\$^{h}$ Argonne National Laboratory, Argonne, IL 60439, USA
\\ $^{i}$ Computational Cosmology Center, Lawrence Berkeley National Laboratory, Berkeley CA 94720, USA
\\$^{j}$ Department of Physics and Atmospheric Science, Dalhousie University, Halifax, NS, B3H 4R2, Canada
\\$^{k}$ Physics Department, McGill University, Montreal, QC H3A 0G4, Canada
\\$^{l}$ Department of Physics, Blackett Laboratory, Imperial College London, London SW7 2AZ, United Kingdom
\\$^{m}$ Department of Astronomy, Pontifica Universidad Catolica, Santiago, Chile
\\$^{n}$ Department of Physics and Astronomy, University of California, Irvine, CA 92697-4575, USA
\\$^{o}$ Center for Astrophysics and Space Astronomy, University of Colorado, Boulder, CO 80309, USA
\\$^{p}$ Department of Physics, Princeton University, Princeton, NJ 08544, USA
\\$^{q}$ Physics Division, Lawrence Berkeley National Laboratory, Berkeley, CA 94720, USA
\\$^{r}$ AstroParticule et Cosmologie, Univ Paris Diderot, CNRS/IN2P3, CEA/Irfu, Obs de Paris, Sorbonne Paris Cit´e, France
\\$^{s}$ Institute of Space and Astronautical Studies (ISAS),Japan Aerospace Exploration Agency (JAXA), Sagamihara, Kanagawa 252-5210, Japan
\\$^{t}$ Observational Cosmology Laboratory, Code 665,NASA Goddard Space Flight Center, Greenbelt, MD 20771, USA
\\$^{u}$ Department of Electrical and Computer Engineering,University of California, San Diego, CA 92093, USA
\\$^{v}$ School of Physics, University of Melbourne, Parkville, VIC 3010, Australia
\\$^{w}$ National Institute for Fusion Science 322-6 Oroshi-cho, Toki City, GIFU Prefecture
\\$^{x}$ SOKENDAI Kamiyamaguchi, Hayama, Miura District, Kanagawa Prefecture 240-0115, Japan
\\$^{y}$ Department of Physics, University of Wisconsin-Madison WI 53706, USA
\\$^{z}$ Department of Physics, Osaka University, Osaka, Japan
\\$^{1}$ Sorbonne Universit\'es, Institut Lagrange de Paris (ILP), 98 bis Boulevard Arago, 75014 Paris, France
}
}


\maketitle

\begin{abstract}

We present an overview of the design and status of the \Pb-2 and the Simons Array experiments. \Pb-2 is a Cosmic Microwave Background polarimetry experiment which aims to characterize the arc-minute angular scale B-mode signal from weak gravitational lensing and search for the degree angular scale B-mode signal from inflationary gravitational waves. The receiver has a 365~mm diameter focal plane cooled to 270~milli-Kelvin. The focal plane is filled with 7,588 dichroic lenslet-antenna coupled polarization sensitive Transition Edge Sensor (TES) bolometric pixels that are sensitive to 95~GHz and 150~GHz bands simultaneously. The TES bolometers are read-out by SQUIDs with 40 channel frequency domain multiplexing. Refractive optical elements are made with high purity alumina to achieve high optical throughput.  The receiver is designed to achieve noise equivalent temperature of 5.8~$\mu$K$_{CMB}\sqrt{s}$ in each frequency band. \Pb-2 will deploy in 2016 in the Atacama desert in Chile. The Simons Array is a project to further increase sensitivity by deploying three \Pb-2 type receivers. The Simons Array will cover 95~GHz, 150~GHz and 220~GHz frequency bands for foreground control. The Simons Array will be able to constrain tensor-to-scalar ratio and sum of neutrino masses to $\sigma(r) = 6\times 10^{-3}$ at $r = 0.1$ and $\sum m_\nu (\sigma =1)$ to 40 meV.

\keywords{Cosmic Microwave Background, Inflation, Gravitational Weak Lensing, Polarization, B-mode}

\end{abstract}

\section{Introduction}
Measurements of the Cosmic Microwave Background (CMB) temperature anisotropy successfully constrained many cosmological parameters.. The CMB is also weakly linearly polarized. The CMB polarization could give tighter constrains on cosmological parameters and open windows for studying fundamental physics. Measurements of even parity polarization pattern, E-mode, of the CMB agree with temperature anisotropy measurements \cite{Beringer:1900zz}. Recently, initial measurements of the odd parity polarization pattern, B-mode, of the CMB were also reported.
\cite{Ade:2014afa,Ade:2014xna,Ade:2013hjl, Ade:2013gez, 0004-637X-811-2-126,Hanson:2013hsb,Keisler:2015hfa}. 

The B-mode polarization has two primary sources. Primordial gravitational waves, if present, would polarize the CMB at degree angular scale \cite{PhysRevLett.78.2054}. Tighter upper limits or detection of the primordial B-mode signal will put constraints on the inflation model and energy level of the inflation potential. Weak gravitational lensing from large scale structures distorts the E-mode pattern to produce small amounts of B-mode polarization pattern \cite{Hu:2001kj}. The B-mode signal from weak gravitational lensing peak around ten arcmin angular scales. Characterization of gravitationally lensed B-mode signal could constrain parameters such as the sum of neutrino masses, evolution of the dark energy equation of state, primordial magnetic fields and cosmic birefringence. Precise characterization of the gravitational lensing B-mode will be important to decouple the lensing signal from the primordial inflationary signal.

Planck's report on the CMB foregrounds suggests that polarized foregrounds such as synchrotron radiation and dust emission need to be carefully subtracted for accurate CMB polarization measurements \cite{Ade:2015tva, refId0}. The \Pb-2 receiver is a highly sensitive receiver with broad frequency coverage for foreground mitigation.

\section{Project Overview}
The \Pb-2 receiver will observe from the James Ax observatory at 5,200 meter altitude in the Chilean Atacama Desert. The site has access to $80\%$ of the sky. This allows cross-correlation with other experiments. The \Pb-2 receiver will be mounted on a telescope with same design as the Huan Tran Telescope (HTT) that is currently observing with the \Pb-1 receiver. The HTT features an offset Gregorian design obeying the Mizuguchi-Dragone condition to minimize instrumental cross-polarization.  The HTT has co-moving baffles to minimize sidelobes. 3.5 meter primary mirror produces a 3.5-arcmin (5.2-arcmin) FWHM beam at 150~GHz (95~GHz). The \Pb-2 receiver will have instantaneous array sensitivity of 5.8~$\mu$K$_{CMB}\sqrt{s}$ in each frequency band. 

The Simons Array is a project to further increase sensitivity by deploying three \Pb-2 type receivers including the \Pb-2 receiver. The first receiver will deploy at 95~GHz and 150~GHz frequencies in 2016. The second receiver will cover 95~GHz and 150~GHz, and the third receiver will cover 150~GHz and 220~GHz bands. The second and third receivers will deploy in 2017. Sensitivity of the Simons Array in its final configuration is 4.1~$\mu$K$_{CMB}\sqrt{s}$ in the 95~GHz band, 3.4~$\mu$K$_{CMB}\sqrt{s}$ in the 150~GHz band and 11.5~$\mu$K$_{CMB}\sqrt{s}$ in the 220~GHz band. The Simons Array will be able to constrain the tensor-to-scalar ratio $r$ to $\sigma(r) = 4\times 10^{-3}$ when considering statistical noise alone, and $\sigma(r) = 6\times 10^{-3}$ at $r = 0.1$ when foregrounds are cleaned \cite{PhysRevD.84.063005}. The Simons Array will also be able to constrain the sum of neutrino masses to 19 meV ($1\sigma$) when considering statistical noise alone, and 40 meV ($1\sigma$) when foreground effect is considered with foreground cleaning by cross-correlation with spectroscopic galaxy surveys.

\section{Instrument}

\begin{figure}
\begin{center}
\includegraphics[width=16cm,keepaspectratio]{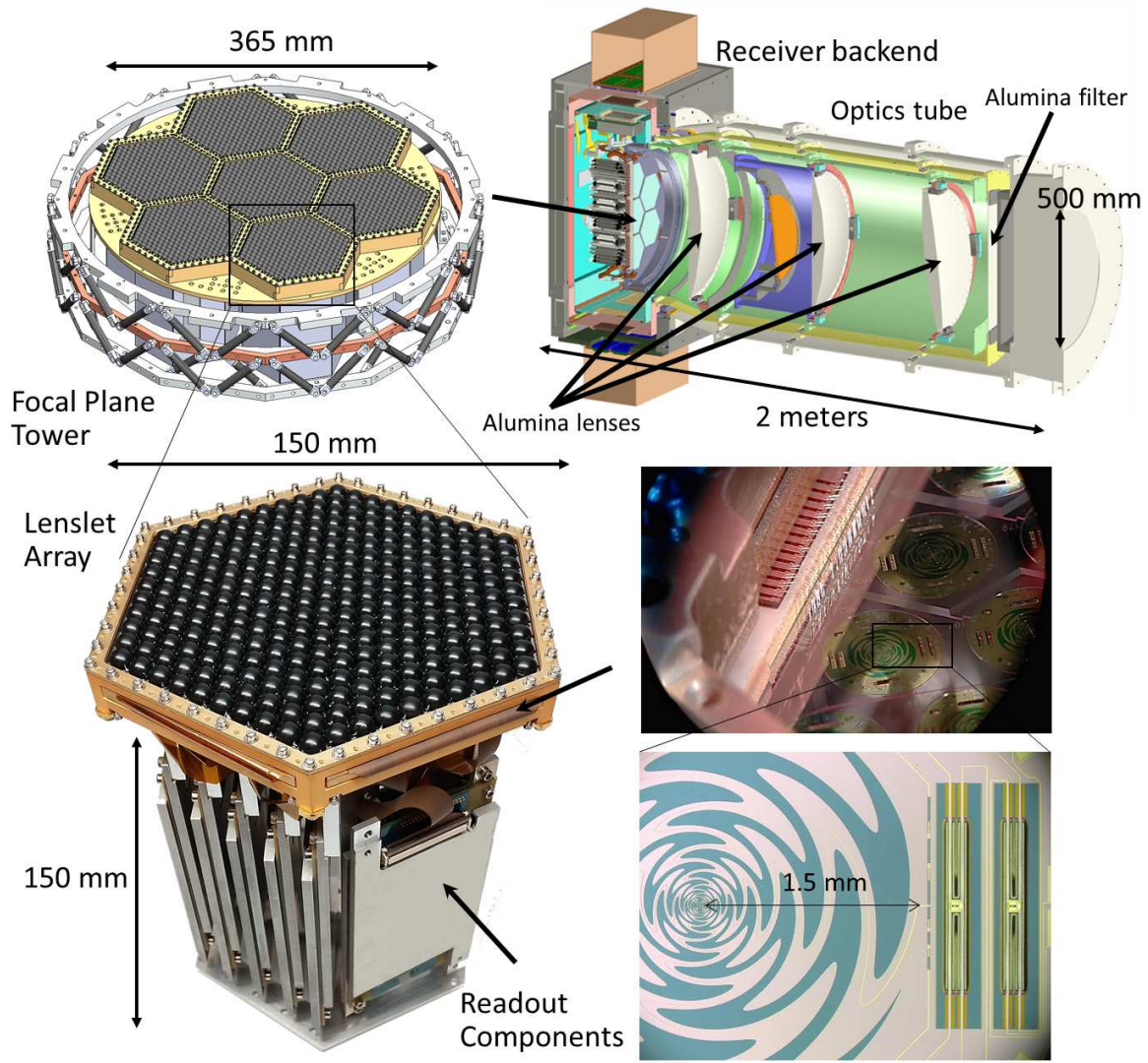}
\end{center}
\caption{(Color online) CAD drawing of the \Pb-2 receiver (upper right), and CAD drawing of the focal plane tower (upper left). Three temperature stages (250~milli-Kelvin, 350~milli-Kelvin, 2~Kelvin) are separated by vespel support structures. Photograph of a detector module (bottom left), which consists of a detector wafer, lenslet wafer, Invar holder, and cryogenic readout electronics. Automated wirebonds have 100 $\mu m$ pitch (bottom right). The Sinusoidal circular structure is a broadband antenna. Large rectangular structures are TES bolometers. The RF diplexer filter is visible between the antenna and the bolometers (bottom right).}
\label{fig:PB2combined}
\end{figure}

A cross-sectional view of the \Pb-2 receiver is shown in Figure~\ref{fig:PB2combined}. Two Cryomech PT415 pulse-tube coolers provide cooling power to the receiver. Annealed 6-N aluminum strips are epoxied to receiver shells to increase thermal conductivity of the receiver. A Chase Cryogenics three-stage helium sorption refrigerator provide the focal plane tower with 2~Kelvin, 350~milli-Kelvin and 270~milli-Kelvin stages.\\
\indent Three reimaging lenses and an infrared filter are fabricated from Nihon Ceratec's $99.9\%$ purity alumina. The high optical index of alumina ($n = 3.1$) minimizes abberation in optics. The alumina has low loss-tangent ($\tan \delta \approx 1\times10^{-4}$), and this keeps receiver efficiency high. High thermal conductivity of the alumina helps with overall cryogenic performance \cite{Inoue:14}. Alumina lenses are anti-reflection coated with two-layer epoxy coating, thermal sprayed ceramic coating and expanded kapton coating \cite{Inoue:14, Rosen:13, JeongLTD16}. The Optical design achieves a strehl ratio greater than 0.90 over entire 365 mm diameter focal plane. The field of view of the \Pb-2 instrument is $4.8^{\circ}$. Optical efficiency of the entire system is 24\% at 95~GHz and 31~\% at 150~GHz.\\
\indent The focal plane is shown in Figure~\ref{fig:PB2combined}. A 365~mm diameter focal plane tower houses seven detector array modules. Each module has 271 dual linear polarized pixels that simultaneously detect CMB radiation in the 95~GHz and 150~GHz bands. Each pixel has a silicon lens coupled broadband sinuous antenna that couples optical signal onto a RF circuit on a wafer. Bandpass filters on the wafer split the signal into two separate bands, then transition edge sensor (TES) bolometers detects the signal \cite{SuzukiLTD15}. Silicon lenslet array is anti-reflection coated with two layers of epoxy based coating \cite{Rosen:13, SiritanasakLTD16}. Readout electronics are assembled behind the detector array for a modular design. \\
\indent TES bolometers are read-out by frequency multiplexed Superconducting Quantum Interference Device (SQUID) amplifiers \cite{HattoriLTD16,:/content/aip/journal/rsi/83/7/10.1063/1.4737629}. Forty channels are frequency multiplexed between 1.6 MHz to 4.2 MHz with logarithmically increasing frequency spacing. Digital Active Nulling technology corrects for phase delay and reduces parasitic inductance from circuit elements between the bias resistor and the SQUID \cite{doi:10.1117/12.2054949}. Superconducting resonators for frequency multiplexing are lithographed on silicon wafers for low loss, high frequency precision and kilo-channel scalability \cite{RotermundLTD16}. We developed superconducting niobium-titanium parallel plate transmission lines for wiring between the milli-Kelvin and 4-Kelvin stages. The low thermal conductivity of niobium-titanium provides thermal isolation, while the high width-to-height ratio of the parallel plate transmission line provides low inductance per length ($\approx\mathrm{1 nH/cm}$) that allows stiff voltage biases of TES 
bolometers.  

\section{Conclusion}
The \Pb-2 and the Simons Array experiment will measure polarization of the CMB with high sensitivity. The \Pb-2 will deploy in 2016, and the Simons Array will fully deploy in 2017.

\begin{acknowledgements}
We acknowledge support from the MEXT Kahenhi grant 21111002, NSF grant AST-0618398, NASA grant NNG06GJ08G, The Simons Foundation, Natural Sciences and Engineering Research Council, Canadian Institute for Advanced Research, Japan Society for the Promotion of Science, and the CONICYT provided invaluable funding and support. Detectors were fabricated at the Berkeley Marvell Nanofabrication laboratory.
\end{acknowledgements}



\end{document}